# High-throughput in-volume processing in glass with isotropic spatial resolutions in three dimensions

**YUANXIN TAN,** [1,2] **ZHAOHUI WANG,** [1,3] **WEI CHU,** [1,5] **YANG LIAO,** [1] **LINGLING QIAO,** [1] **AND YA CHENG**[1,4,*]

[1]*State Key Laboratory of High Field Laser Physics, Shanghai Institute of Optics and Fine Mechanics, Chinese Academy of Sciences, Shanghai 201800, China*
[2]*School of Physical Science and Technology, Shanghai Tech University, Shanghai 200031, China*
[3]*Luoyang Institute of Electro-Optical Equipment, AVIC, Luoyang 471000, China*
[4]*State Key Laboratory of Precision Spectroscopy, East Normal University, Shanghai 200062, China*
[5]*chuwei0818@qq.com*
[*]*ya.cheng@siom.ac.cn*

**Abstract:** We report on fabrication of three dimensional (3D) microstructures in glass with isotropic spatial resolutions. To achieve high throughput fabrication, we expand the focal spot size with a low-numerical-aperture lens, which naturally results in a degraded axial resolution. We solve the problem with simultaneous spatial temporal focusing which leads to an isotropic laser-affected volume with a spatial resolution of ~100 μm.

## 1. Introduction

Femtosecond laser provides a unique tool for processing inside transparent materials [1-2]. The enabling approach is three dimensional (3D) femtosecond laser direct writing (FLDW) which is conducted by scanning tightly focused spots in various transparent materials including glass, crystals, polymers, bio-tissues, etc to produce arbitrary 3D structures [3-9]. Interaction of the focused femtosecond laser pulses with transparent materials leads to modifications of the materials properties ranging from optical refractive index, density, and Young's module to compositional distribution and chemical stability. Importantly, the space-selective modifications enabled by the 3D FLDW provides a spatial resolution comparable to or even exceed that allowed by diffraction limit, thanks to the nonlinear dependence of the modifications in materials on the peak intensity of the femtosecond laser pulses as well as the suppressed thermal diffusion with the ultrashort pulse durations [10,11]. Therefore, 3D FLDW offers very high spatial resolution to construct fine 3D structures through two-photon polymerization or internal processing of glasses or crystals.

One of the challenging issues in 3D FLDW is its relatively low throughput as compared to the conventional photolithography, because it relies on scan of the tiny focal spots produced by high numerical aperture (NA) objective lenses. For some 3D processing applications such as building mechanical molds, high throughput is highly in demand to bring down the fabrication cost at a reasonably sacrificed spatial resolution. However, if one attempts to accomplish this merely by using focal lenses of low NAs, the focal spot will become highly asymmetric, as the focal depth increases quadratically with the increasing NA, whereas the transverse size of the focal spot only increases linearly with the increasing NA [12]. This causes a severe issue in achieving high throughput 3D FLDW which is in favor of using low-NA focal lenses whereas maintaining a nearly isotropic resolution in the modification by the focused femtosecond laser pulses.

In this work, we attempt to overcome the above-mentioned problem with a simultaneous spatial temporal focusing (SSTF) scheme [13-15]. The working mechanism of SSTF is to first spatially chirp the incident pulse before the focal lens, forming an array of beamlets at different frequencies. This leads to the elongation of the pulse duration due to the reduction of the bandwidth of each beamlet. After passing through the focal lens, all of the beamlets will be recombined at the focal plane, where the original broadband spectrum of the incident femtosecond pulse is restored to give rise to the shortest pulses. In such a manner, the peak intensity of the spatiotemporally focused pulses is strongly localized near the geometrical focus, i.e., the longitudinal resolution is improved in the interaction of the spatiotemporally focused pulses with transparent materials. In the first implementation of the SSTF scheme in 3D FLDW, F. He et al has demonstrated an isotropic fabrication resolution of a few micron using an objective lens of NA=0.46 [13]. In the current work, we demonstrate high throughput FLDW of complex 3D structures in glass with an isotropic resolution of ~100 μm. Although our 3D structure is produced in glass, this technique can generally be extended to 3D processing of other types of materials such as two-photon polymerization or ablation of bio-tissue.

## 2. Experimental

The glass material used in the experiment is a well-known photosensitive glass, Foturan, which is composed of a lithium aluminosilicate doped with trace amounts of silver and cerium [16]. During the exposure to the femtosecond laser, free electrons are generated by multiphoton ionization to reduce the silver ions to silver atoms. By a subsequent heat treatment, silver atoms diffuse to form nanoparticles. Due to the plasmon resonance scattering of the silver nanoparticles, the laser modified area appears brown [17,18]. With this glass, the 3D intensity distribution of femtosecond pulses at the focus can be recorded and inspected under the microscope [15]. In this sense, Foturan glass can be regarded as a 3D photographic recording medium. In this experiment, commercially available Foturan glass was cut into 5 mm × 5 mm × 1.5 mm coupons with the all six sides polished to record the 3D spatial profile of simultaneously spatio-temporally focused spot.

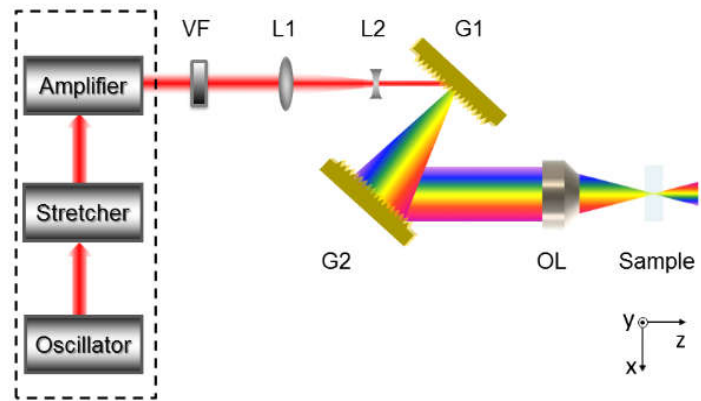

Fig. 1 Schematic illustration of the experimental setup. VF: variable neutral density filter. G1-2: gratings (1500 l/mm). L1, L2: lenses of different focal lengths which are described in the main text. OL: objective lens. The configuration of XYZ axes are indicated in the inset.

Figure 1 schematically illustrates the experimental setup. The femtosecond laser system (Libra-HE, Coherent Inc.) used in this experiment consists of a Ti: sapphire laser oscillator and amplifier, and a grating-based stretcher and compressor that delivers 3.5 mJ, 50 fs pulses with a spectral bandwidth of ∼26 nm centered at 800 nm wavelength at 1 kHz repetition rate. In the SSTF, the compressor was bypassed and the diameter of the amplified laser beam was first reduced to ∼0.55 mm ($1/e^2$) using a telescope system consisting of a convex lens ($f$ = 800 mm) and a concave lens ($f$ = −50 mm). The beam were then directed through the single-pass grating compressor, consisting of two $\sigma$ = 1500 grooves/mm gratings, blazed for the incident angle of 53°. The distance between the two gratings was adjusted to be ~730 mm to compensate for the temporal dispersion of pulses. After being dispersed by the grating pair, the laser beam was measured to be ~40 mm ($1/e^2$) along the x-axis and ~0.55 mm ($1/e^2$) along the y-axis. The spatially dispersed laser pulses were then focused into the glass sample using an objective lens (Leica, 2 ×, NA =0.35) with a focal length of $f$ = 40 mm. However, since we had reduced our beam diameter to ~0.55 mm ($1/e^2$), the back aperture of the lens was severely under filled. The effective NA estimated in this circumstance was only 0.013. In our experiment, the average power of laser beam was controlled using a variable neutral density (ND) filter VF. The glass samples can be arbitrarily translated three dimensionally by a PC-controlled XYZ stage with a resolution of 1 μm. The machining process was monitored by a CCD camera.

## 3. Results and discussions

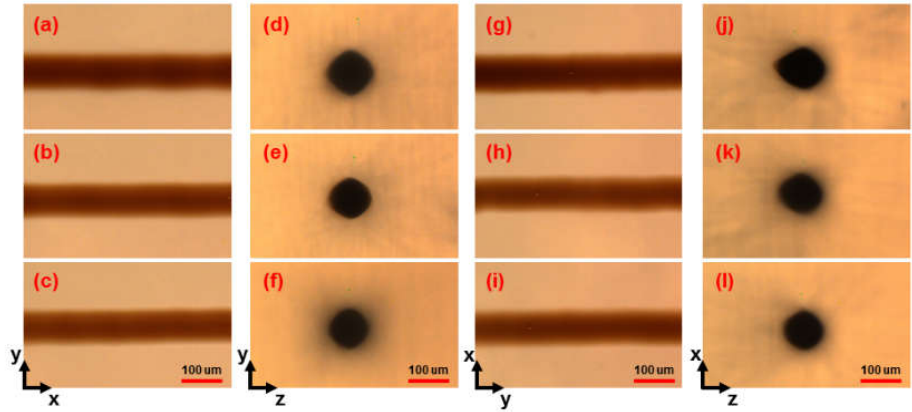

Fig. 2 (a-c, g-i) Top-view and (d-f, j-l) cross-sectional-view optical micrographs of several lines inscribed in Foturan glass along x axis (a-f) and z axis (g-i) at different laser powers. The laser powers are 80 mW in the first row (a, d, g, j), 70 mW in the second row (b, e, h, k) and 60 mW in the third row (c, f, i, l).

Figure 2 shows various lines inscribed in Foturan glass along different scan directions at different laser intensities for individually examining the spatial resolutions along XYZ axes. We inscribed parallel lines in both the X- and Y-directions in Foturan glass samples with the SSTF scheme. The writing speed was fixed at 5 mm/s and the average laser power varied from 60 to 80 mW (measured before the grating pair). Under the above irradiation conditions, no visible modifications of the glass could be observed under the optical microscope after the femtosecond laser irradiation. Then the coupon was subjected to a programmed heat treatment. The temperature was first ramped from room temperature to 500 °C at 5 °C/min and held at 500 °C for 1 hr; then it was raised to 605 °C at 5 °C/min and again held for 1 hr. After the samples were naturally cooled down to room temperature, the laser modified area appears brown. We polished the samples and examined the modified areas using the optical microscope.

Figure 2(a-c) show the optical micrographs of lines oriented along x direction written by setting the average laser power at 80 mW, 70 mW and 60 mW, respectively. The widths of lines are all approximately 85 um. We examined the cross sections of these lines by cutting the samples and polishing their facets. Figure 2(d-f) show the cross sectional profiles of the lines, which are almost round-shaped. Next we fabricated three lines along y direction with all the fabrication parameters unchanged. The optical micrographs of the lines are shown in Fig. 2 (g-i) whilst the corresponding cross sectional profiles are shown in Figure 2(j-l). It can be observed clearly that all the lines oriented in both the x and y directions exhibit a nearly circular cross section with a diameter of 85 μm. Particularly, when the average laser power was set at 60 mW, a perfectly isotropic resolution in all XYZ axes was achieved.

To further examine the maximum laser direct writing speed achievable at the isotropic fabrication resolution of ~85 μm, we fabricated four parallel lines in Foturan glass samples at various speeds ranging from 4 mm/s to 8 mm/s. The femtosecond laser power was fixed at 60 mW. Figure 3(a-d) show the top-view micrographs of the lines inscribed by setting the writing speeds at 4 mm/s, 5 mm/s, 6 mm/s and 8 mm/s, respectively. Figure 3(e-h) are the cross-sectional profiles of the lines in Fig. 3(a-d). It can be seen that the cross sections of the lines all appear symmetric at the different writing speeds, and the diameters of the lines are all approximately 85 μm. However, the lines become obviously inhomogeneous when the speed is higher than 6 mm/s (e.g., see Fig. 3(c) and (d)), which should be caused by insufficient overlap between consecutive pulses during the laser direct writing. It should be noted that our femtosecond laser operates at 1 kHz repetition rate. Therefore, higher writing speeds can be expected at higher repetition rates with SSTF scheme from the results in Fig. 3. The lines are homogeneous and continuous when the writing speed is under 5 mm/s, which ensures a high fabrication quality as shown in Fig. 3(a) and (b).

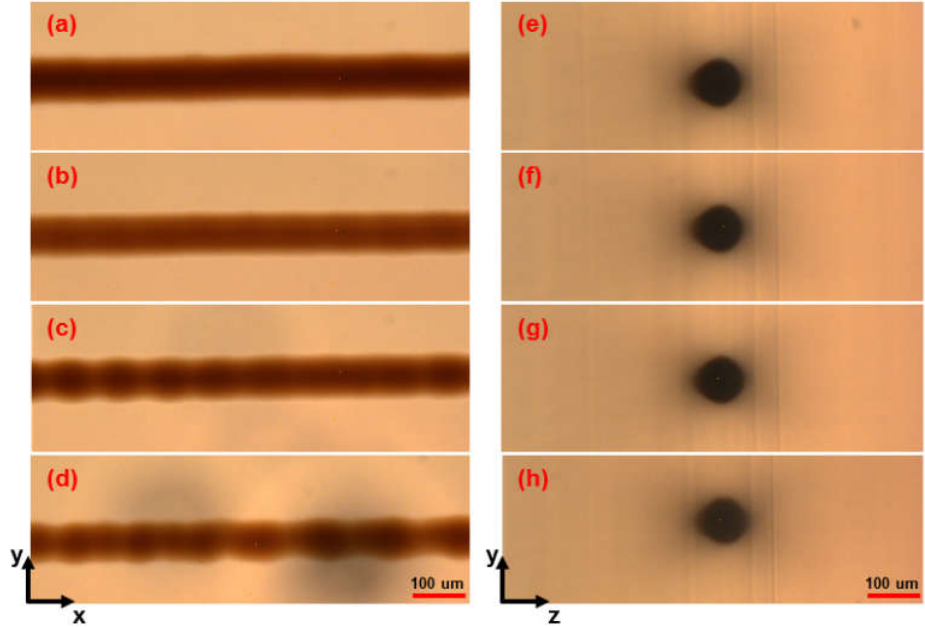

Fig. 3 (a-d) Top-view and (e-h) cross-sectional view optical micrographs of four lines written at different speeds with SSTF scheme. The speeds are 4 mm/s in (a, e), 5 mm/s in (b, f), 6 mm/s in (c, g), and 8 mm/s in (d, h).

For comparison, we also performed the experiments of writing straight lines in different focal conditions and at different powers using the conventional focusing scheme. In this case, the amplified pulses were first compressed by passing through the double-pass compressor and then focused by the same objective lens used in spatiotemporal focusing scheme. The writing speed was fixed at 100 μm/s. Figures 4(a)-(c) show the top-view micrographs of the lines inscribed in Foturan glass samples by focusing the full-size femtosecond laser beam with the conventional focusing scheme. The average laser powers were set at 0.5 mW, 0.75 mW, 1 mW (measured before the objective lens) in Fig. (a)-(c), respectively. The obtained line-widths were measured to be 12.2 μm, 14.6 μm, and 15 μm in (a)-(c), respectively. Figure 4(e)-(g) show the cross sectional profiles of the lines in Fig. 3(a-c), showing that the depth of the lines in Fig. (a)-(c) are 164 μm, 210 μm, and 233 μm, respectively. The results prove that with the conventional focusing, the focal spot produced by a low-NA lens is highly asymmetric, leading to seriously degraded axial resolution.

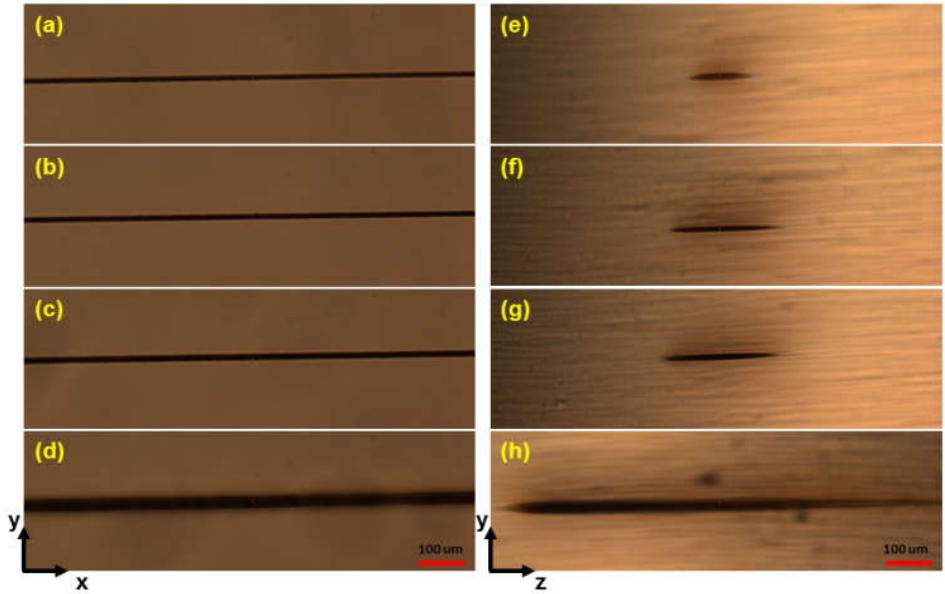

Fig. 4 (a-d) Top-view and (e-h) cross-sectional view optical micrographs of three lines written at different speeds with conventional focusing scheme. A full-size beam was used in (a-c) and (e-f), whereas a beam of reduced diameter of 0.55 mm (1/e2) was used in (d) and (h). The average powers of femtosecond laser are 0.5 mW in (a, e), 0.75 mW in (b, f), 1mW in (c, g) and 3 mW (d, h).

Furthermore, we performed the laser writing of lines in Foturan glass with the conventional focusing scheme in which the effective NA maintains the same as that used in SSTF scheme. For this purpose, the diameter of the laser beam after the double-pass compressor was reduced to 0.55 mm ($1/e^2$) with the same telescope system as shown in Fig. 1. The beam was then focused into the glass sample with the same objective lens. The writing speed was fixed at 100 μm/s and the average laser power was set to be 3 mW (measured before objective lens). The higher laser power was chosen because of the looser focusing condition. The top-view and cross-sectional view optical micrographs of the line are shown in Fig. 4 (d) and (h), respectively. The lateral linewidth was measured to be 32 μm, whereas the depth of the lines along the optical axis of the lens was measured to be 854 μm. The results clearly show that it is impossible to obtain large voxel sizes of 3D isotropic resolution by use of the conventional focusing scheme.

To experimentally demonstrate the unique capability of high-throughput fabrication of large scale 3D structures with SSTF scheme, we inscribed a 3D model of China Pavilion of EXPO 2010 in Foturan glass, as shown by its digital-camera-captured image in Fig. 5. In this demonstration, the experimental parameters are the same as that in Fig. 2(c), which provides the most isotropic resolution in XYZ directions. The length, width, and height of the model are 3.5 mm, 3.5 mm, and 1.5 mm, respectively. It only took 6 minutes to complete the writing of the whole structure despite that our experiment employs a femtosecond laser at a relatively low repetition rate of 1 kHz.

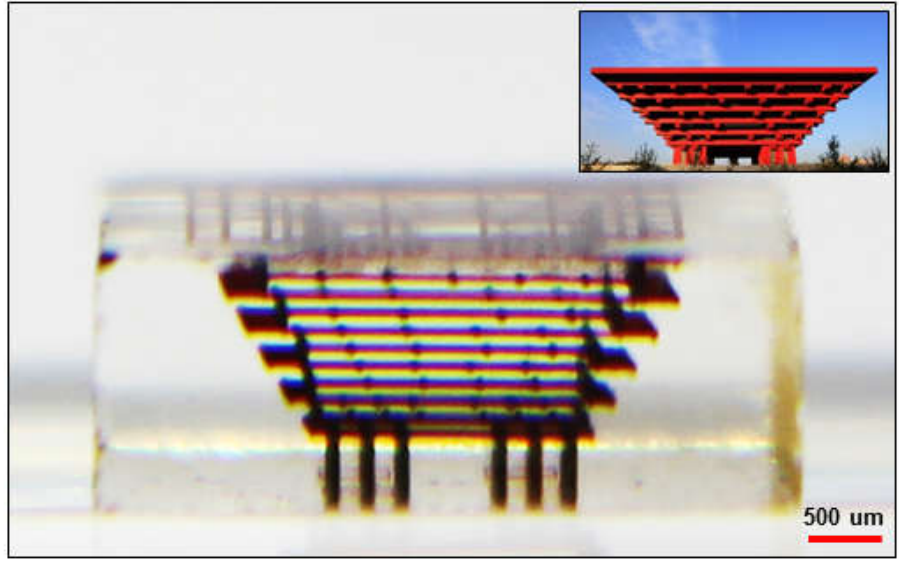

Fig. 5 Image of the China Pavilion written in Foturan glass captured by digital camera

## 4. Conclusion

To conclude, we have shown the potential of using the SSTF scheme for high-throughput 3D femtosecond laser fabrication. The voxel size achieved in our experimental is approximately 85 μm × 85 μm × 85 μm, which is almost 6 orders of magnitude larger than the focal spot produced by a high NA objective lens. This leads to a one-million-fold enhancement of the fabrication efficiency. It is flexible to tune the voxel size by controlling the spatial chirp of the incident pulse at the back aperture of the lens based on a tradeoff between the fabrication resolution and the production yield. Our technique can also be used in two-photon polymerization for producing bio-compatible 3D structures at higher production rates.

## Acknowledgments

This work was supported by the National Basic Research Program of China (Grant No. 2014CB921300), National Natural Science Foundation of China (Nos. 61327902, 11104294, 61275205, 61405220, 61221064 and 11304330), and the Youth Innovation Promotion Association of Chinese Academy of Sciences.